\documentclass[aps,pre,twocolumn,showpacs,showkeys,floatfix]{revtex4}

\usepackage{graphicx}
\usepackage{dcolumn}
\usepackage{bm}

\newcommand{\grad}{\mbox{grad}}
\renewcommand{\div}{\mbox{div}}
\newcommand{\pdiffl}[2]{\frac{\partial #1}{\partial #2}}

\begin{document}


\title{On the Properties of Plastic Ablators in Laser-Driven Material Dynamics
   Experiments}

\date{December 6, 2007; revised April 5, 2008 and June 3, 2008
  -- UCRL-JRNL-236641}

\author{Damian C. Swift}
\email{dswift@llnl.gov}
\affiliation{%
   CMELS-MSTD, Lawrence Livermore National Laboratory,
   7000 East Avenue, Livermore, California 94550, USA
}

\author{Richard G. Kraus}
\affiliation{%
   Department of Physics, Cavendish Laboratory, University of Cambridge,
   JJ~Thomson Avenue, Cambridge CB3~0HE, UK
}


%
%
%
%
%

\begin{abstract}
Radiation hydrodynamics simulations were used to study the effect of
plastic ablators in laser-driven shock experiments.
The sensitivity to composition and equation of state was found to be
5-10\%\ in ablation pressure.
As was found for metals, 
a laser pulse of constant irradiance gave a pressure history
which decreased by several percent per nanosecond.
The pressure history could be made more constant by adjusting the
irradiance history.
The impedance mismatch with the sample gave an increase $o(100\%)$ 
in the pressure transmitted into the sample,
for a reduction of several tens of percent in the duration of the peak load
applied to the sample,
and structured the release history by adding a release step to a pressure
close to the ablation pressure.
Algebraic relations were found between the laser pulse duration, the ablator
thickness, and the duration of the
peak pressure applied to the sample, involving quantities calculated
from the equations of state of the ablator and sample using shock dynamics.
\end{abstract}

\pacs{07.35.+k, 52.38.Mf, 47.40.Nm, 79.20.Ds}
\keywords{shock physics, laser ablation}

\maketitle

\section{Introduction}
Studies of the properties and response of matter under extreme conditions
rely increasingly on lasers to induce dynamic loading by ablation.
Compared with the canonical technique of impact-induced shocks,
laser ablation is more convenient for studying phenomena on shorter
time scales (nanoseconds rather than microseconds), for investigating the
properties of single crystals and the basic microstructural processes,
and for developing new diagnostics.
However, laser ablation has potential disadvantages including uncertainties in
loading history, spatial variations in irradiance from speckles or the
overall intensity envelope, and the possibility of heating the sample
under investigation by hot electrons or Bremsstrahlung x-rays.
A thorough understanding of the processes of ablative loading is necessary
to assure that the response of the sample can be distinguished from
phenomena associated with the ablation process.

In many experiments, dynamic loading can be induced by ablation of the
sample material itself.
Previous studies have demonstrated that spatial variations and preheat were
negligible, for ablatively-generated shocks from $\sim$100\,MPa to 1\,TPa
in elements \cite{Swift_elements_04} and compounds \cite{Swift_alloys_04},
and for ramp compression \cite{Swift_lice_05}.
In other experiments, it is highly desirable to use the laser energy to
ablate a well-controlled layer of a different material, in contact with the
sample.
Advantages of a separate ablator include the ability to restrict 
multi-electron-volt ablation temperatures to elements of low atomic number
irrespective of the sample material, reducing the risk of generating hard
x-rays; smoothing of small-scale spatial variations by beam dispersion,
radiation transport, and mechanical equilibration within the ablation plasma;
and the ability to use a single ablator of known plasma properties for
experiments on less well-understood sample materials.
A current interest in dynamic loading experiments is the use of
x-ray diffraction to study the sample material in the compressed state
\cite{Remington06}.
With ramp compression, the material of highest compression is closest to
the ablation surface, so it is desirable to pass the x-rays in and out through
the ablator and the ablation plume.
It is thus particularly desirable to restrict the ablator to
elements of low atomic number, dissimilar to the sample.
When the sample material is ablated directly, heat conduction from the plasma 
and re-condensation of ablated material at the end of the laser pulse can 
lead to material with a locally different microstructure 
\cite{Peralta05}, an added complication when interpreting the microstructure
of recovered samples.
A dissimilar ablator acts as a palliative in this case.
By appropriate choice of material, an ablator may also increase the efficiency
of conversion of laser energy to pressure in the sample.

An ablator layer can however complicate dynamic loading experiments.
The time at which the load is first applied to the sample depends on
the ablator thickness and the laser irradiance.
The mechanical impedance mismatch between the ablator and the sample causes
wave interactions, leading to multiple waves in the sample which must be
discriminated from the response of the sample.
If the composition and properties of the ablator material are not adequately
known, it introduces additional uncertainty in the experiments.

Plastics are attractive ablator materials as they are stable to handle and 
readily available 
with compositions of low atomic number, such as polystyrene (C$_8$H$_8$)$_n$ 
and polyparaxylylene-N (parylene-N)
(C$_8$H$_8$)$_n$, and polyethene (C$_2$H$_4$)$_n$. 
Deposition and coating techniques exist which can produce layers of
uniform and well-characterized thickness \cite{Powell66,Mattox98}.
Polystyrene and
parylene-N ablators have been used as a coating for glass spheres in
inertial confinement fusion experiments.
Previous studies have investigated the properties of the ablation plasma
including two dimensional effects in planar targets
\cite{Grun81}, and have investigated the generation and mitigation of preheat
in the sample \cite{Delettrez90,Colvin05}.
Significant discrepancies have been identified between pressures inferred
from experiments and predicted in
radiation hydrodynamics simulations \cite{Delettrez90}.
Parylene layers have been used in 
material dynamics experiments at lower irradiances,
again with some discrepancy between simulation and experiment
\cite{Hawreliak06}.

Here we present studies of the behavior of plastic ablators under conditions
relevant to shock physics and material dynamics experiments, including 
sensitivities to uncertainties in the composition or 
equation of state of the ablator, and wave interactions caused by the
impedance mismatch with the sample. 
The effect of wave interactions is presented in a form convenient for
the design of laser-driven shock experiments.

\section{Radiation hydrodynamics simulations}
The response of a plastic layer to intense laser irradiance,
and the effect of the impedance mismatch between the plastic and a denser
sample material behind, were investigated using numerical simulations.
Laser-matter interaction is non-linear and time-dependent,
so spatially-resolved radiation hydrodynamics is needed to integrate
the coupled  radiation transport and continuum dynamics equations forward
in time.
In the regime of interest, laser ablation and dynamic loading can be simulated
accurately by assuming three temperature hydrodynamics (ions, electrons,
and radiation), including thermal conduction and radiation diffusion,
and calculating the absorption of the laser energy in the expanding
plasma cloud through the electrical conductivity \cite{Swift_elements_04}.
Thermal conduction and radiation diffusion are necessary only near the
ablated surface of the plastic: the bulk of the ablator and the sample behind
are affected only by the hydrodynamics.

Material dynamics experiments are designed to apply a simple loading history
to the sample: ideally close to one-dimensional in space, i.e. with a 
time-dependence that is independent of position over a region of the
sample surface.
Experiments are usually designed so that edge effects, two- or three-dimensional
dependencies in the loading history, do not affect some portion of the sample
for long enough for the one-dimensional response to be measured.
We therefore use one-dimensional radiation hydrodynamics, and estimate the
size of the region affected by the edges.

Radiation hydrodynamics is the simultaneous solution of the continuum
dynamics, heat conduction, and radiation transport equations
\cite{Mihalas84,LASNEX}.
Using the Lagrangian frame of reference, in which time derivatives are
expressed along characteristics moving locally with the material,
the equations of continuum dynamics are
\begin{eqnarray}
\frac{D\rho}{Dt} & = & -\rho\div\,u \\
\frac{Du}{Dt} & = & -\frac 1\rho\grad\,p \\
\frac{De}{Dt} & = & -\frac p\rho \div\,u,
\end{eqnarray}
where $t$ is time, $\rho$ the mass density, $u$ the particle velocity,
$p$ the pressure, and $e$ the specific internal energy.
These equations are closed by a mechanical equation of state (EOS) $p(\rho,e)$.
Each equation may also have a source term; here, heat conduction and
radiation transport provide source terms in the velocity and energy equations.
Heat conduction and radiation transport involve temperatures
for the ions $T_i$, electrons $T_e$, and radiation field $T_r$.
In the regime of interest, laser and thermal radiation couples principally
to the electrons, which equilibrate with the ions according to a 
compression- and temperature-dependent rate $\tau_{ei}(\rho,T)$.
A thermal EOS relates $e$ to $T_i$; in practice we used an EOS of the
form $\{p,e\}(\rho,T_i)$, using $T_i$ as one of the state parameters and
calculating $e(T)$ and $\dot T_i(\dot e)$ when needed in the hydrodynamics.
The equation of heat conduction is
\begin{equation}
\pdiffl Tt = \div\left[\kappa(\rho,T_i)\grad\,T\right],
\end{equation}
where $\kappa(\rho,T_i)$ is the thermal conductivity.
Radiation transport was calculated for an equilibrium radiation distribution
(represented by a single radiation temperature field $T_r$)
The equation of single-temperature radiation diffusion is
\begin{equation}
\pdiffl Ut = -p_r\div\,u-\div F+4\pi\bar\alpha_a B_P - c(\bar\alpha_a+\bar\alpha_s)U,
\end{equation}
where $U\equiv 4\sigma T_r^4/c$,
$\bar\alpha_a$ and $\bar\alpha_s$ are frequency-averaged absorption and
scattering coefficients,
$B_P$ is the integrated power in the Planck distribution at $T_r$,
the radiation flux
\begin{equation}
F = -\frac{4\sigma \div T_r^4}{3\bar\kappa_R\rho+|\div T_r^4|/T_r^4},
\end{equation}
$\sigma$ is the Stefan-Boltzmann constant, and $c$ is the speed of light.
The absorption coefficient was approximated as the Rosseland mean opacity, 
and the emission coefficient as the Planck mean opacity.
The transport and deposition of incident laser energy was calculated
as the attenuation of a free-propagating flux, using the electrical
conductivity to calculate the attenuation, with energy deposited as heating of
the electrons.

In our context, radiation hydrodynamics is an initial value problem,
where the material fields $\rho$ and $e$ are specified over a region $R$ 
at some time $t_0$,
time-dependent boundary conditions are specified 
for the continuum $p_{\partial R}(t)$ or $u_{\partial R}(t)$
and radiation flux $I_{\partial R}(t)$,
and the continuum and radiation fields are integrated for $t > t_0$.

Simulations were performed using the
HYADES radiation hydrocode, version 01.05.11 \cite{HYADES}.
This hydrocode used a one-dimensional (1D)
Lagrangian finite-difference discretization 
of the material and leapfrog time integration,
with shock waves stabilized using artificial viscosity.
The radiation flux limiter was set to 0.03 of the free stream value -
a common choice for simulations of this type \cite{Dendy93}.

\section{Properties of plastic ablators}
The ablation properties of plastic coatings were investigated by
radiation hydrodynamics simulations as described above,
to predict the relationship between irradiance and pressure, ablation rates,
and the sensitivity to composition of the plastic.

The properties of the plastic were represented through models for its
pressure-volume-energy equation of state (EOS), its conductivities,
and its opacity.
Wide-ranging EOS were taken from the SESAME database \cite{SESAME}.
Opacities for radiation diffusion were also taken from SESAME.
Conductivities for laser deposition and heat conduction were
calculated using the Thomas-Fermi ionization model
\cite{HYADES,Zeldovich66};
this was found previously to be accurate for direct drive
shock simulations on samples of a wide range of atomic numbers
\cite{Swift_elements_04}.
A significant difficulty with parylene-N is the only EOS model available was
constructed by density scaling from an EOS for parylene-C, which was fitted
to shock Hugoniot data at 300-600\,GPa with an uncertainty $\sim$10\%\ 
\cite{Young07,Rothman02}.
In parylene-C, one H is substituted by Cl, and density scaling will not 
necessarily capture the different chemical behavior.
To investigate the uncertainty in pressures predicted in ablative loading
experiments, the sensitivity to EOS and opacity was investigated by
performing simulations for parylene-C and parylene-D, in which an additional
H atom is substituted by Cl,
and also for polystyrene, which has the same composition
as parylene-N though a different molecular structure.
The molecular structure should become unimportant once the
plastic has been converted to plasma.
The models for parylene-D (SESAME tables 7770 and 17770) \cite{Johnson-Lyon91} 
were developed earlier and have been used more commonly in
radiation hydrodynamics simulations, so these models were used for the
studies of general ablation behavior described next.

To allow a fine spatial resolution of the surface to be ablated,
without requiring the same resolution in the bulk of the plastic where
it remains solid, the material was discretized with 
geometrically-expanding cells.
The expansion factor between successive cells was
around 1-5\%: small enough to avoid inaccuracy in the hydrodynamic equations.
The smallest cell, at the surface where ablation starts,
was chosen to have an initial size in the few nanometer range.
Numerical convergence with respect to spatial resolution 
was tested by performing simulations with 200 or 300
zones, with an expansion ratio of 2.5\%\ or 2.0\%\ respectively,
giving zone widths of around 9.0 or 2.6\,nm at the ablation surface.
The ablation pressure was around 5\%\ higher in the finer-mesh simulations,
indicating an adequate level of mesh convergence.
The finer mesh was used for all subsequent simulations.

For the ablation rate and pressure study, the plastic was chosen to be
50\,$\mu$m thick, and the laser pulse was 10\,ns long with constant irradiance.
To avoid numerical oscillations, the irradiance was taken to rise from and fall
to zero over 0.1\,ns, which is typical of the laser systems used for the
material dynamics experiments of interest here.
The critical surface of the plasma was predicted to remain within a few 
micrometers of the initial position of the ablator, so the plasma plume
should be essentially 1D.
Release and cooling at late times would be faster than predicted using
1D simulations, because of lateral expansion.

\begin{figure}
\begin{center}\includegraphics[scale=0.72]{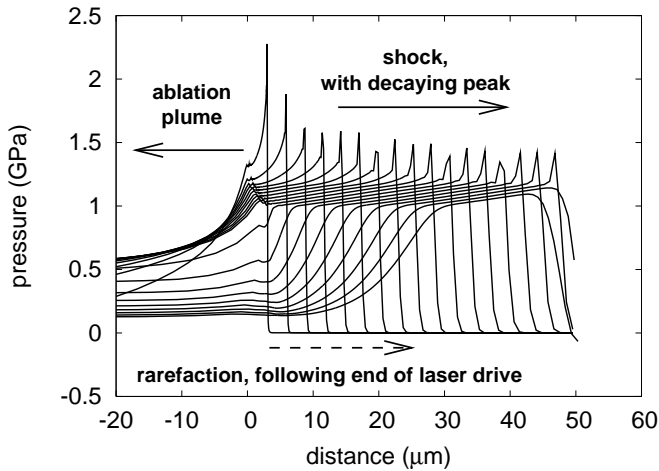}\end{center}
\caption{Predicted pressure profiles in parylene-N at intervals of 1\,ns,
   for an irradiance of 0.1\,PW/m$^2$ applied for 10\,ns.
   The undisturbed material extends from 0 to 50\,$\mu$m,
   and the laser drive impinges from the left.}
\label{fig:ch1f}
\end{figure}

In the simulations, the surface of the plastic was heated to plasma
conditions rapidly after the start of the laser pulse.
The plasma then absorbed the incident laser energy, 
forming a nearly steady state plume 
close to the plastic.
The plasma pressure and reaction to plasma flow induced an elevated pressure
in the remaining condensed-phase plastic, driving a shock wave.
As was found previously \cite{Swift_elements_04,Swift_alloys_04},
a laser pulse of constant irradiance did not induce a constant driving
pressure in the condensed region.
Instead, there was an initial spike of pressure tens of percent higher, decaying
rapidly to a more constant sustained pressure.
This sustained pressure typically decreased by a few percent during the 
laser pulse.
The EOS used exhibited simple, concave behavior, so the decaying pressure in the
spike and the plateau was able to catch up and erode the higher pressure
states ahead, reducing the shock pressure monotonically as it propagated 
through the material.\footnote{%
In an optimized shock-loading experiment, the pulse shape of the laser is
adjusted to induce a more constant pressure history.
Constant irradiance histories are considered here for simplicity.
}
After the end of the pulse, the ablation pressure fell rapidly,
and a rarefaction wave propagated into the shocked material,
releasing the pressure.
For the thickness and pulse duration chosen, and over the irradiance range
considered, the shock reached the rear surface of the plastic before the
rarefaction caught up, so a second rarefaction wave from the rear surface
also propagated into the material.
(Fig.~\ref{fig:ch1f})

\begin{figure}
\begin{center}\includegraphics[scale=0.72]{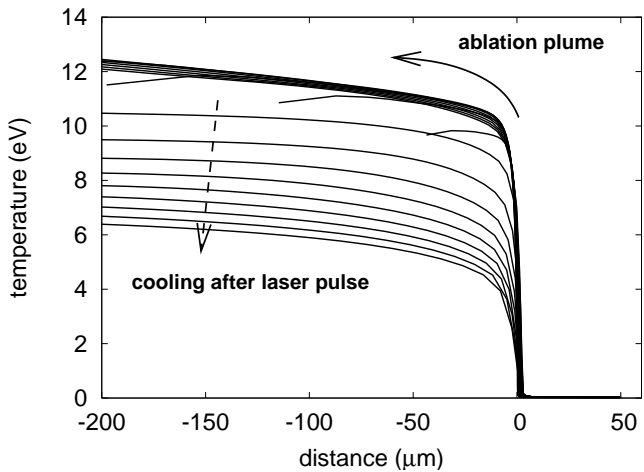}\end{center}
\caption{Predicted profiles of ion temperature in parylene-N
   at intervals of 1\,ns,
   for an irradiance of 0.1\,PW/m$^2$ applied for 10\,ns.
   The undisturbed material extends from 0 to 50\,$\mu$m,
   and the laser drive impinges from the left.}
\label{fig:ch1ft}
\end{figure}

Temperatures in the ablation plume were predicted to be of order
10-100\,eV.
The mass density and time scales were low enough that conductive
heating of the condensed-phase ablator was not significant.
(Fig.~\ref{fig:ch1ft})

The relationship between irradiance and ablation pressure was investigated
by performing simulations with the same 10\,ns laser pulse duration,
and varying the irradiance between 0.1 and 100\,PW/m$^2$.\footnote{%
Irradiances are commonly expressed in the non-SI unit of 
GW/cm$^2$; 1\,PW/m$^2$=100\,GW/cm$^2$.
}
The sustained pressure $p$ following the initial peak was represented
well by a power law in irradiance $I$:
\begin{equation}
p (\mbox{GPa}) = 8.614 \left[I (\mbox{PW/m$^2$})\right]^{0.833}
\end{equation}
The fitting uncertainty in the parameters is 1.5\%,
and the uncertainty in the irradiance-pressure values is around 5\%,
because of temporal variations in the pressure.
This result is consistent with previous studies on elemental metals
\cite{Swift_elements_04}, though the exponent here is larger,
and the irradiance-pressure relation is better reproduced by 
the power law fit despite covering a wider range of irradiance.
The exponent is consistent with experimental and calculational results
obtained for polystyrene \cite{Grun81}, and the prefactor is similar to
the value implied by these results.
The exponent is also consistent with more recent simulation results
for parylene \cite{Colvin05}.

\begin{figure}
\begin{center}\includegraphics[scale=0.72]{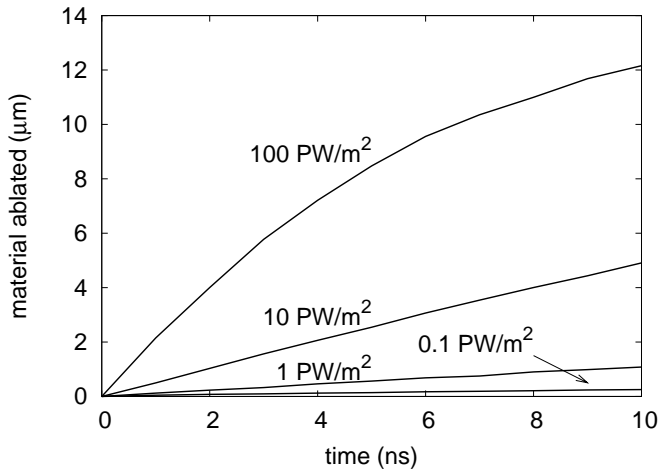}\end{center}
\caption{Penetration of ablation surface during the laser pulse,
   for different irradiances.}
\label{fig:ablratecmp}
\end{figure}

In this regime of irradiance, laser and thermal radiation penetrates the 
material much more slowly than the shock wave passes through,
so there is a region of compressed material heated only by the passage
of the shock between the ablation surface and the undisturbed material
ahead of the shock.
It is important to know the rate at which material is ablated, to ensure that
the ablator is thick enough for a laser pulse of given irradiance and duration.
The ablation rate was extracted from the simulations by observing the time of
first outward motion of points in the material.
For convenience in choosing a thickness of ablator, a Lagrangian ablation rate
was calculated, i.e. with respect to the original position of the material.
The ablation-supported shock compresses material before it is ablated,
so the ablation rate, with respect to the instantaneously compressed 
material, is lower than the Lagrangian rate.
For laser irradiances from 0.1 and 10\,PW/m$^2$,
the ablation rates were fitted well by a straight line, indicating that a
constant irradiance produces a constant ablation rate.
At an irradiance of 100\,PW/m$^2$, the ablation rate decreased significantly
over the 10\,ns of the laser pulse.
Ablation was predicted to continue at a similar rate 
for around 20\%\ of the duration of the
laser pulse after its end, because of retained heat in the ablation plume.
After this time, the ablation rate decreased rapidly
(Fig.~\ref{fig:ablratecmp}).
The individual ablation rates $u_A$ were reproduced accurately by the relation
\begin{equation}
u_A (\mbox{km/s})=0.108 \left[I (\mbox{PW/m$^2$})\right]^{0.651},
\end{equation}
valid for $I$ between 0.1 and 100\,PW/m$^2$.
Above 10\,PW/m$^2$, this relation ceases to be valid over the full 10\,ns
investigated: at 100\,PW/m$^2$ it is valid for the first 1\,ns.
The fitting uncertainty in the parameters is 0.5\%.
Note that 1\,km/s=1\,$\mu$m/ns.
This result is similar to the results reported for the ablation of polystyrene
\cite{Grun81}, though our simulations predict a larger prefactor.
In the previous results, the depth of material ablated was inferred indirectly
from measurements of the ablation plasma.
\footnote{%
In the previous work \cite{Grun81}, `ablation velocity' referred to the speed of
the expanding plasma, and not to the speed of the ablation front through the 
compressed ablator.
}

\begin{figure}
\begin{center}\includegraphics[scale=0.72]{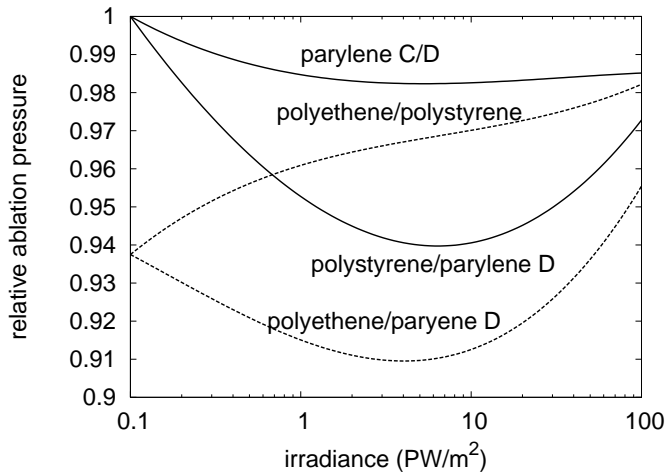}\end{center}
\caption{Sustained ablation pressure predicted for different plastics,
   with respect to the pressures calculated for parylene-D or polystyrene.}
\label{fig:pablcmp}
\end{figure}

There is significant uncertainty in the EOS of matter, particularly
in the warm dense matter regime of the ablation plume.
The uncertainty is correspondingly greater for a material for which no
specific EOS has been developed, as discussed above.
It was found previously \cite{Swift_elements_04,Swift_alloys_04}
that pressures induced in the condensed region were fairly insensitive
to the details of the plasma EOS.
The sensitivity was investigated here by performing equivalent 
sets of simulations
using the EOS and opacity for parylene-C from the SESAME library:
(tables 7771 and 17620) \cite{Johnson-Lyon91}, 
and also polystyrene (tables 7590 and 17593) \cite{Barnes-Lindstrom76}.
The two parylene models gave the same ablation pressure to within 1-2\%,
and the shock speeds were almost identical.
The polystyrene model gave a smaller initial pressure spike
and a sustained pressure that was up to several percent lower,
but with less of a decrease over the pulse duration
and hence a slower decay as it propagated through the plastic.
The shock speed in polystyrene was significantly different:
slower at low pressures and faster at high pressures.
These variations are all small compared with typical uncertainties in
mean irradiance in laser ablation experiments.
(Fig.~\ref{fig:pablcmp}.)

The sensitivity to ablator material was investigated further by performing
equivalent simulations for polyethene (tables 7171 and 17171)
\cite{Dowell82}.
The ablation pressures were 5-10\%\ lower than for parylene-D, but within 5\%\
of polystyrene.
The pressure profile was generally similar to that of polystyrene.
(Fig.~\ref{fig:pablcmp}.)

\begin{figure}
\begin{center}\includegraphics[scale=0.72]{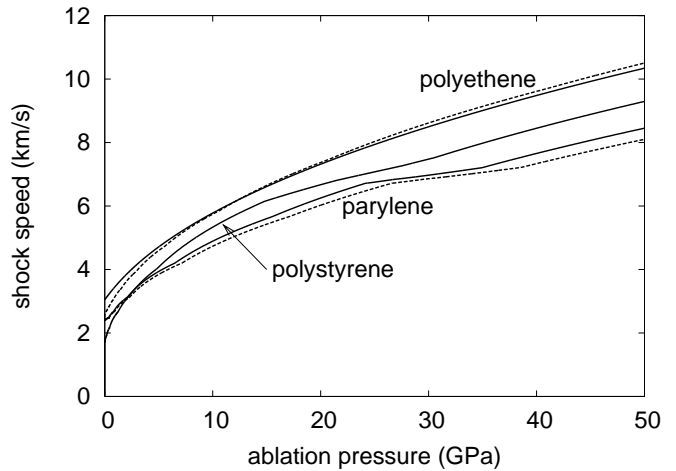}\end{center}
\caption{Shock speed in the ablator as a function of sustained ablation
   pressure, for different ablator compositions and equations of state.
   For polyethene, the solid and dashed lines are Steinberg's analytical
   equation of state and SESAME table 7171 respectively.
   For parylene, solid and dashed lines are for parylene-C and D
   respectively.}
\label{fig:hugpus}
\end{figure}

For x-ray diffraction experiments, the timing of the x-ray pulse with 
respect to the shock-loading of the sample is particularly important.
The effect of composition and EOS was significantly more sensitive than the
ablation pressure
(Fig.~\ref{fig:hugpus}).

\begin{figure}
\begin{center}\includegraphics[scale=0.72]{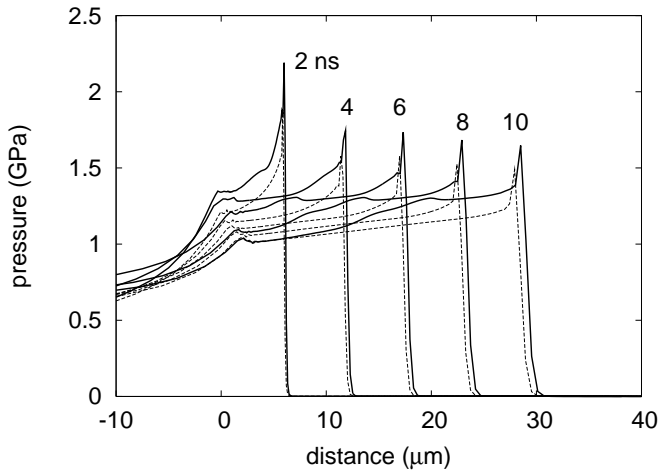}\end{center}
\caption{Predicted pressure profiles in parylene-N at intervals of 2\,ns,
   with (solid lines) and without (dashed lines) a flashing of 100\,nm Al
   on the surface of the plastic.
   The laser irradiance was 0.1\,PW/m$^2$ applied for 10\,ns.
   The undisturbed material extends from 0 to 50\,$\mu$m,
   and the laser drive impinges from the left.}
\label{fig:alch1cmp}
\end{figure}

Experimentally, it has been found that some laser light may be transmitted
through the plastic coating before the opaque plasma sheath forms,
inducing earlier ablation of the sample material behind by a small amount
of the laser energy.
This early transmission was not predicted in the radiation hydrodynamics
simulations, which (as is common) did not include an adequate model of the
line opacity and breakdown of the plastic.
A palliative is to deposit a thin, metal flashing on the surface of the plastic
to prevent early transmission.
A typical flashing is 100\,nm of Al, though other metals have been used
\cite{Delettrez90}.
The Al layer potentially has an effect on the ablation pressure, apart from
the palliative effect.
This effect was predicted by performing simulations in which the first
100\,nm of the plastic was replaced by Al.
At 0.1\,PW/m$^2$, the presence of the Al increased the sustained ablation 
pressure by 10-15\%\ and made it more constant,
but introduced slight wave reverberations in the
compressed ablator (Fig.~\ref{fig:alch1cmp}).
The difference was smaller at 1\,PW/m$^2$,
and negligible at 10\,PW/m$^2$ and above.

There is also a potential effect from x-ray preheating of the sample,
induced by K-shell radiation from the ablator or flashing.
X-ray preheating is very sensitive to intensity variations in the laser
beam, as x-ray yield is highly non-linear in laser irradiance
\cite{Workman01}.
The fluence of preheating x-rays is therefore very sensitive to the
beam quality of the specific laser system,
including any phase plate used to reduce large-scale intensity variations.
The degree of x-ray heating also depends on the opacity of the sample material.
Experimental evidence and theoretical predictions for x-ray preheating in
such experiments is mixed: simulations have indicated that preheating may be
significant in our regime of interest \cite{Colvin05}, but whereas experiments
have shown little or no preheating 
\cite{Grun81,Swift_elements_04,Swift_alloys_04,Peralta05}.
While preheating is a potential concern when the initial temperature of
the sample is important, it has a much smaller effect on wave interactions
caused by the presence of the ablator, so we do not consider it further in the
present work.

At the end of the laser pulse, the pressure at the ablation surface drops 
rapidly as the ablation plume expands and cools.
Until two-dimensional expansion takes effect, the pressure gradient in
the ablation plume is inversely proportional to the pulse length,
so the rate at which the pressure falls is inversely proportional to the
pulse length.

\section{Laser pulse shaping for constant ablation pressure}
The ablation pressure can be made constant in time by adjusting the
temporal shape of the laser pulse.
Many, though not all, large laser systems allow the power history to be
controlled, so a desired temporal shape can be delivered to some finite level
of precision.
On lasers with large numbers of beams but a simple shape from each 
(such as a constant power), a smooth temporal shape can be 
approximated by altering the energy and relative timing of the beams.

\begin{figure}
\begin{center}\includegraphics[scale=0.72]{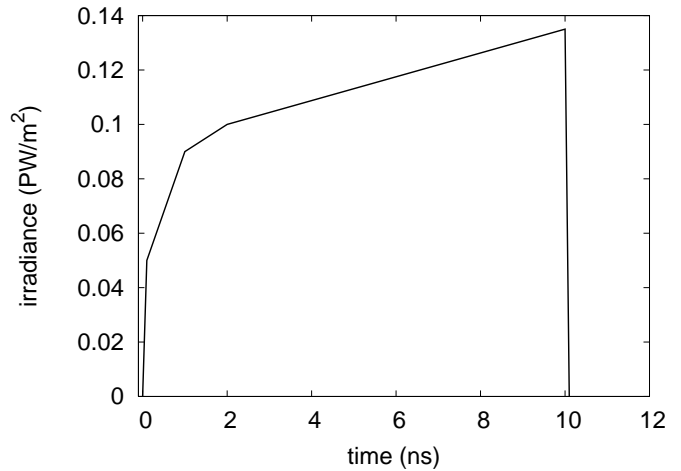}\end{center}
\caption{Example laser irradiance history designed to give a more constant shock
   pressure than from a constant irradiance of 0.1\,PW/m$^2$.}
\label{fig:pulseshape}
\end{figure}

The principle of pulse shaping for constant ablation pressure was demonstrated
for parylene-N ablators by performing simulations in which the laser
irradiance was adjusted.
These simulations were performed for a laser pulse where
the time-dependent irradiance was represented in tabular form,
with table entries at the start and end of the pulse of equal value for
a constant irradiance.
The procedure followed was first to adjust the irradiance at the end of the
pulse, giving a linear ramp starting at the initial irradiance,
and then to add extra points between the first and last and
change their values to remove shorter time scale variations in pressure.
The sustained ablation pressure was made constant by ramping the irradiance
up in time, and the initial pressure spike was removed by reducing
the irradiance early in time and following a more gentle approach to the
main part of the pulse.
This procedure was demonstrated previously for direct ablation of Be
\cite{Luo05}.
Calculationally, it was possible to make the induced pressure constant to any 
desired precision by adjusting the laser irradiance history
(Fig.~\ref{fig:pulseshape}).
For irradiances from 0.1 to 10\,PW/m$^2$, the ramp was 10-30\%\ of the mean
irradiance.

\section{Wave interactions with the sample}
When the ablator is in contact with a sample, the impedance mismatch between
the ablator and the sample lead to wave interactions which change the
pressure history in the ablator and may lead to multiple waves in the
sample.
In most cases of interest, the sample has a higher shock impedance than the
plastic ablator, so the ablation-induced shock reflects a shock from the
interface between the ablator and the sample.

\begin{figure}
\begin{center}\includegraphics[scale=0.55]{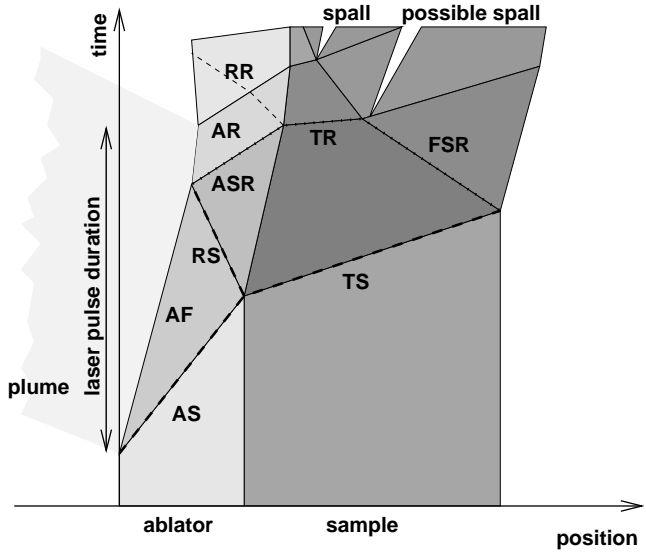}\end{center}
\caption{Schematic of wave interactions in position-time space
   for laser loading in one dimension
   with an ablator of lower impedance than the sample.
   Diagram shows the ablation-driven shock (AS),
   transmitted shock (TS) in the sample,
   reflected shock (RS) in the ablator;
   ablation surface release (ASR), transmitted release (TR),
   reflected release (RR), and
   free surface release (FSR);
   ablation release (AR); ablation front (AF).
   Darker shades represent material of higher mass density.
   For simplicity, release waves are shown as lines rather than fans.}
\label{fig:ablwavext}
\end{figure}

\begin{figure}
\begin{center}\includegraphics[scale=0.65]{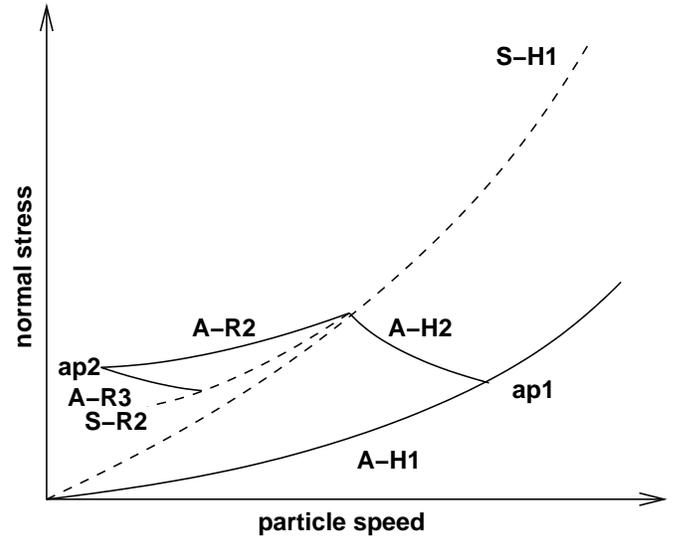}\end{center}
\caption{Schematic of wave interactions in particle speed-normal stress space
   for laser loading in one dimension
   with an ablator of lower impedance than the sample.
   Loci for the ablator (A) and sample (S) are solid and dashed respectively.
   The principal shock Hugoniots are denoted H1.
   Ablation induces a state ap1 on the principal Hugoniot of the ablator.
   Because of the higher impedance of the sample,
   the state induced is the intersection of the sample's principal Hugoniot
   with the secondary Hugoniot H2 of the ablator.
   When the reflected shock reaches the ablation surface,
   a release wave R2 propagates through the ablator,
   releasing to the ablation pressure of the re-shocked ablator material,
   ap2.
   When this release wave reaches the sample,
   the ablator material must release further (R3) to intersect the
   release adiabat R2 of the sample.
   This intersection gives the next state induced in the sample.}
\label{fig:ablwaveupp}
\end{figure}

In this discussion of wave interactions, we consider a simplified
situation where the laser irradiance history has been adjusted to
give a constant drive pressure in the ablator,
which will induce a constant initial shock pressure in the sample.
Early in the laser pulse, an ablation-driven shock is induced,
driving a transmitted shock in the sample and a reflected
shock in the ablator.
If the laser pulse is long enough,
when the reflected shock reaches the ablation surface,
double-shocked ablator material is then ablated, resulting in an ablation
pressure that is slightly higher than before, but generally lower
than the double-shocked pressure.
The resulting ablation surface release wave propagates through the 
ablator and is transmitted into the sample,
reflecting a further weak release wave back through the ablator.
When the transmitted shock reaches the free surface of the sample,
a release wave propagates backward into the sample.
When the laser pulse ends, a strong release wave propagates through any
remaining ablator and forward into the sample.
Where the strong release waves interact, the sample material is subjected to
tension, and spall may occur.
(Figs~\ref{fig:ablwavext} and \ref{fig:ablwaveupp}.)

\begin{figure}
\begin{center}\includegraphics[scale=0.72]{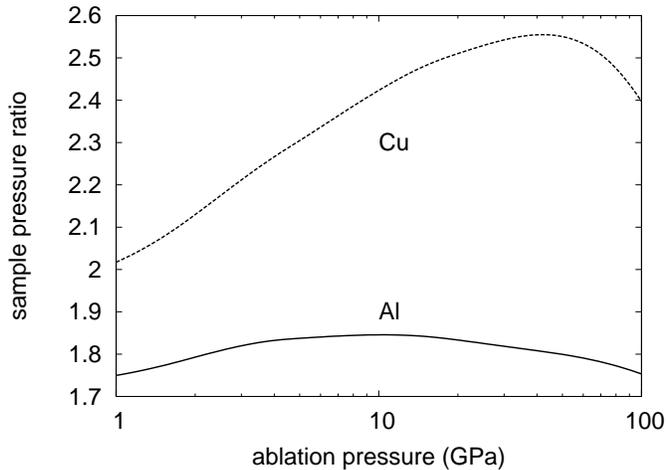}\end{center}
\caption{Effect of shock impedance mismatch for plastic ablators
   driving shocks into Al and Cu samples.
   The sample pressure ratio is the pressure in the sample divided by the
   ablation pressure.}
\label{fig:impedmismatch}
\end{figure}

The impedance mismatch between most metals and plastics 
such as parylene and polyethene means that a given ablation pressure in the
plastic induces a considerably higher pressure in a metal sample.
The pressure enhancement was estimated for the plastic EOS used above
and published EOS for Al and Cu, by finding the intersection between the 
secondary shock Hugoniot for a given initial shock state in the plastic 
and the principal Hugoniot in the metal in pressure-particle speed space.
The EOS of Al was as above; that for Cu was SESAME table 3330.
Shock Hugoniots and their intersections were found numerically
\cite{Swift_genhug07}.
Repeating this procedure for a range of ablation shock pressures,
relations were found between the pressure in the plastic and the pressure
in the metal (Fig.~\ref{fig:impedmismatch}).


\begin{figure}
\begin{center}\includegraphics[scale=0.72]{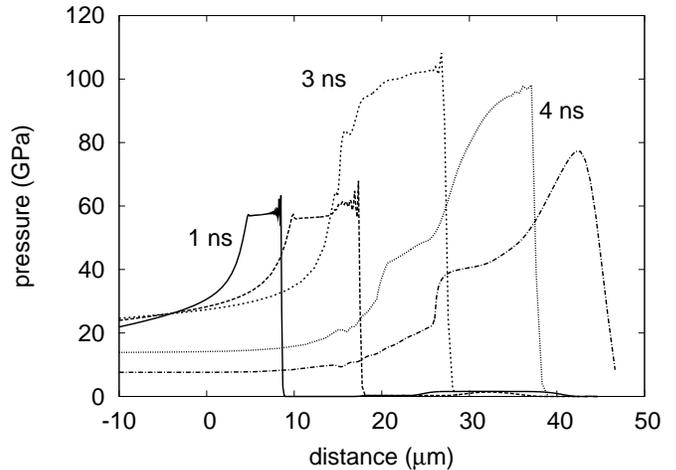}\end{center}
\caption{Pressure profiles at intervals of 1\,ns for an Al sample
   coated with 20\,$\mu$m of parylene-N ablator, flashed with 100\,nm of Al,
   and driven with a laser pulse of 10\,PW/m$^2$ for 3\,ns.
   Before the start of the laser drive, the ablator
   extended from 0 to 20\,$\mu$m, and the sample from 20 to 45\,$\mu$m.
   The profile at 1\,ns shows the shock in the ablator.
   At 3\,ns, the shock has been transmitted into the sample,
   almost doubling the pressure but exaggerating its deviation from flatness.
   At 4\,ns, the shock has almost reached the free surface of the sample,
   its peak has decayed and become less flat, and the ablation surface release
   is evident as a tilted step between 40 and 50\,GPa on release.
   By 5\,ns, the shock has reached the free surface of the sample and started
   to release.}
\label{fig:sampleprofcmpAl}
\end{figure}

\begin{figure}
\begin{center}\includegraphics[scale=0.72]{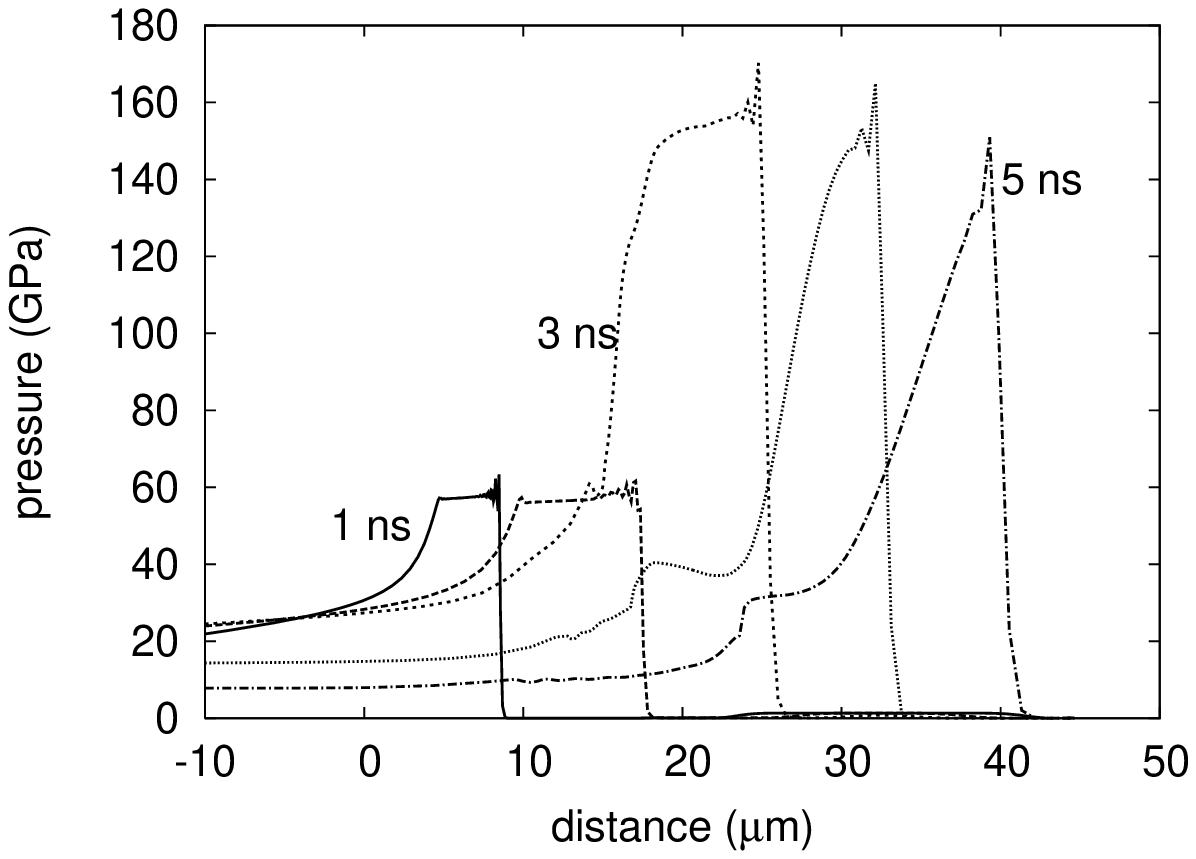}\end{center}
\caption{Pressure profiles at intervals of 1\,ns for a Cu sample
   coated with 20\,$\mu$m of parylene-N ablator, flashed with 100\,nm of Al,
   and driven with a laser pulse of 10\,PW/m$^2$ for 3\,ns.
   Before the start of the laser drive, the ablator
   extended from 0 to 20\,$\mu$m, and the sample from 20 to 45\,$\mu$m.
   The profile at 1\,ns shows the shock in the ablator.
   At 3\,ns, the shock has been transmitted into the sample,
   almost trebling the pressure but exaggerating its deviation from flatness.
   At 5\,ns, the shock has almost reached the free surface of the sample,
   and the ablation surface release has caught up with the shock and made it
   triangular.}
\label{fig:sampleprofcmpCu}
\end{figure}

\begin{figure}
\begin{center}\includegraphics[scale=0.72]{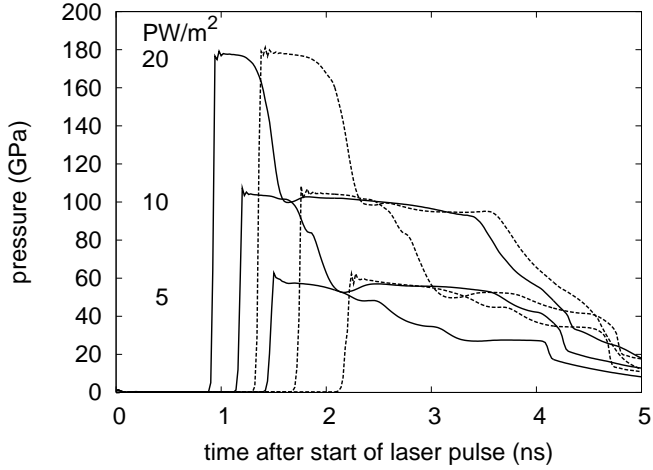}\end{center}
\caption{Effect of ablator thickness on the pressure history applied
   to Al samples.
   Simulations for 10\,$\mu$m (solid) and 20\,$\mu$m (dashed) ablators,
   with a 3\,ns drive pulse of 5, 10, and 20\,PW/m$^2$.}
\label{fig:ablthickcmp}
\end{figure}

Radiation hydrodynamics simulations were performed to predict the 
integrated effect of ablative loading via a plastic ablator on the
loading history experienced by the sample.
As for the ablator-only simulations, a geometrically expanding 
spatial discretization was used.
The sample was chosen to be 25\,$\mu$m thick.
Simulations were performed for ablators 10 and 20\,$\mu$m thick,
with Al flashing.
The laser pulse was taken to be 3\,ns long, with constant irradiance,
which are specifications often used for material dynamics experiments
at the Janus laser \cite{Hawreliak06}.
With a laser pulse of constant irradiance, the pressure was predicted to
decrease with time during the pulse, though the compression varied less.
Simulations were performed for irradiances of 5, 10, 20, and 50\,PW/m$^2$.
The peak pressure in the sample was in agreement with the impedance mismatch
calculations, and decayed as the shock propagated through the sample.
With a thicker ablator, the pressure in the sample was more constant.
(Figs~\ref{fig:sampleprofcmpAl} to \ref{fig:ablthickcmp}.)

\begin{figure}
\begin{center}\includegraphics[scale=0.72]{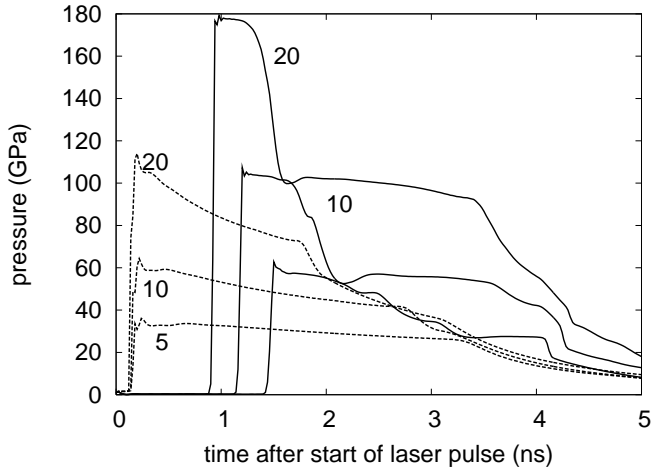}\end{center}
\caption{Comparison between pressure induced by laser ablation in Al 
   samples by a laser pulse 3\,ns long of constant irradiance,
   with (solid) and without (dashed) a parylene-N ablator.
   Irradiances are given in PW/m$^2$.
   The pressure history was calculated at the ablator-sample interface
   when the ablator was present,
   and at a Lagrangian position 1\,$\mu$m inside the sample
   when there was no ablator.}
\label{fig:pablcmpAl}
\end{figure}

\begin{figure}
\begin{center}\includegraphics[scale=0.72]{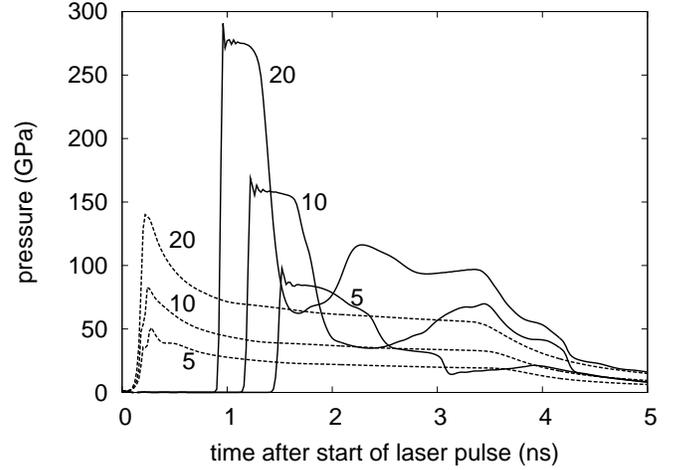}\end{center}
\caption{Comparison between pressure induced by laser ablation in Cu 
   samples by a laser pulse 3\,ns long of constant irradiance,
   with (solid) and without (dashed) a parylene-N ablator.
   Irradiances are given in PW/m$^2$.
   The pressure history was calculated at the ablator-sample interface
   when the ablator was present,
   and at a Lagrangian position 1\,$\mu$m inside the sample
   when there was no ablator.}
\label{fig:pablcmpCu}
\end{figure}

For comparison, simulations of ablative loading of uncoated samples
were performed.  For a given irradiance, the pressure induced in the
sample was significantly lower when ablated directly, though the
duration of the period of high pressure was greater.
Ablation of plastic was predicted to generate a slightly lower pressure
than ablation of the metal sample, but the large impedance mismatch
more than compensated for the reduced ablation pressure.
With an ablator, the pressure fell more slowly at the end of the laser pulse.
With an ablator, the pressure varied less during the initial shock,
with a constant laser irradiance.
The inclusion of the low density ablator increased the coupling of
laser energy to the sample.
(Figs~\ref{fig:pablcmpAl} and \ref{fig:pablcmpCu}.)

One way to think of the function of the ablator is that it converts laser energy
into kinetic and potential energy in the moving, compressed
ablator, which is then partly transferred to the sample.
For a given irradiance, laser pulse length, and sample material, 
there is an optimum thickness for any ablator 
to maximize the duration of the peak shock pressure applied to the sample.
The optimum thickness is such that the laser pulse ends when the reflected
shock reaches the ablation surface.
When the ablation rate is a constant in time, this relationship can usefully
be expressed as an effective speed: 
microns of ablator per nanoseconds of laser pulse;
this is a useful quantity when designing ablation experiments.
The optimum relationship between thickness and pulse length depends on the sample material
too, because the impedance mismatch affects the pressure of the reflected shock
and therefore its transit time.
It should be emphasized that this `ablator drive speed' does not represent any individual
physical wave speed: it is a composite of the initial ablator shock speed, the
speed of the reflected shock through the pre-compressed ablator,
and the ablation rate.
If the laser pulse is the optimum length for the ablator thickness, or longer,
then the duration of the higher pressure drive experienced by the sample at the
interface with the ablator can be related to the duration of the laser pulse,
as a load duration factor.
If the laser pulse is longer than the optimum for a given ablator thickness,
the principal pressure pulse applied to the sample does
not become longer, but the sample pressure releases to the ablation pressure,
and this lower pressure is sustained for longer.
There is a minimum laser pulse duration for the peak pressure to reach the 
sample.
If the laser pulse is longer than this, but
shorter than the ablator thickness optimum,
the duration of the pressure pulse at the sample varies proportionately from
zero to the maximum.

The initial mass density of the ablator is $\rho_0$.
The ablation-driven shock travels at a speed $s_a$ and compresses the ablator
to a mass density $\rho_a$, ablating material at a rate $u_A$ with respect
to the uncompressed ablator (the quantity calculated above)
or $u_a=u_A\rho_0/\rho_a$ of the compressed ablator.
The shock reflected from the sample back through the ablator travels
at a speed $s_r$ and compresses the ablator further to a mass density $\rho_r$.
In this state, the speed of sound is $c_r$: this is the speed of the head
of the ablation surface rarefaction.
For an ablator of initial, uncompressed thickness $l_0$,
the transit time of the initial ablation shock is $\tau_s=l_0/s_a$: this is the
time that loading starts in the sample after the start of the laser pulse.
The transit time of the reflected shock is
\begin{equation}
\tau_r=l_0\frac{\rho_0/\rho_a-u_a/s_a}{s_r+u_a}.
\end{equation}
The optimum laser pulse length is $\tau_s+\tau_r$, 
so the ablator drive speed is
\begin{equation}
u_d=\frac{s_a\rho_a(s_r+u_a)}{s_a\rho_0+s_r\rho_a}.
\end{equation}
The residual thickness of ablator when the reflected shock reaches the
ablation surface is
\begin{equation}
l_r=l_0\left[\rho_0/\rho_a-u_a\left(\tau_s+\tau_r\right)\right]\frac{\rho_a}{\rho_r},
\end{equation}
so the transit time of the ablation surface release is 
\begin{equation}
\tau_a=l_r/c_r=\frac{l_0s_r\left(\rho_0s_a-\rho_au_a\right)}{\rho_rc_rs_a\left(s_r+u_a\right)}.
\end{equation}
The time for which the initial pressure is applied to the sample is thus
\begin{equation}
\tau_l=\tau_r+\tau_a
=\frac{l_0(\rho_0s_a-\rho_au_a)(\rho_as_r+\rho_rc_r)}{\rho_a\rho_rc_rs_a(s_r+u_a)}
\end{equation}
so the load duration factor is 
\begin{equation}
f_l=\frac{\tau_l}{\tau_s+\tau_r}
=\frac{(\rho_0s_a-\rho_au_a)(\rho_as_r+\rho_rc_r)}{\rho_rc_r(\rho_0s_a+\rho_as_r)}.
\end{equation}

If the sound speed in the ablation shocked state is $c_a$,
traveling through the compressed ablator,
the minimum pulse length for the peak pressure to reach the sample
can be found by calculating the point in the ablator at which the head of
a release wave at the end of the laser pulse would catch up with the
ablation shock.
The minimum time is proportional to the sample thickness, so an effective
`catch-up speed' can be defined for a given ablation pressure:
\begin{equation}
u_c=\frac{s_a c_a}{\rho_a c_a-\rho_0s_a},
\end{equation}
where $s_a$, $c_a$, and $\rho_a$ are all in the principal Hugoniot state for
the ablator at the ablation pressure.
For a given laser pulse duration $\tau_d$, the maximum depth in the ablator
before the ablation shock starts to decay is then $u_c\tau_d$.
This calculation can also be used to calculate the maximum sample thickness
for a supported shock in experiments where the sample itself is ablated
\cite{Swift_maxthick_01}.
For an ablator of given thickness $l$, the minimum laser pulse duration
$\tau_{\mbox{min}}=l/u_c$.
For laser pulse durations between this and 
$\tau_{\mbox{max}}\equiv\tau_a+\tau_r$,
the duration of the peak pressure pulse applied to the sample is
\begin{equation}
\tau_s=f_l\tau_{\mbox{max}}\frac{\tau_d-\tau_{\mbox{min}}}{\tau_{\mbox{max}}-\tau_{\mbox{min}}}.
\end{equation}

\begin{figure}
\begin{center}\includegraphics[scale=0.60]{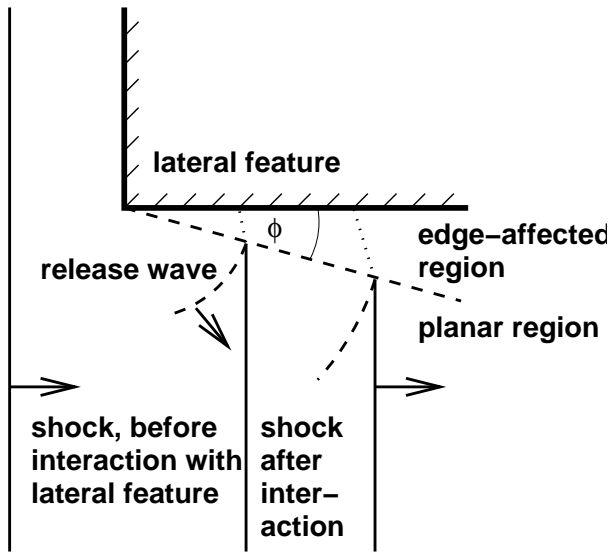}\end{center}
\caption{Schematic of edge release eroding the shocked region.
   The lateral feature causing edge release is shown here as an
   interface (corner) with a material of lower shock impedance.
   The shock wave is shown at three positions: one just before
   interaction with the feature, where it is planar over the field
   of view, and two at different times after the interaction,
   showing the erosion of the shock by the rarefaction from the edge.}
\label{fig:latrel}
\end{figure}

As a planar loading or unloading wave propagates through a sample,
release waves from the edge will usually erode the shocked region.\footnote{%
   The exception is if `edge faking' using an angled boundary with a material
   of higher shock impedance is present to give a perfect impedance match,
   which is difficult in general, and may reflect a shock or release if not
   designed correctly.
}
The head of the release wave initiated by any lateral variation
in the material or 
loading conditions is an expanding circle with respect to the shocked
material (Fig.~\ref{fig:latrel}): a cylinder or torus in two-dimensional plane
or axisymmetric geometry respectively.
The lateral variation in a projectile impact experiment is the edge of
the projectile on the impact surface;
for laser ablation it is the edge of the focal spot -- though
unlike a projectile impact, the pressure does not release
to zero because the ablation of laterally-released material
may induce a pressure almost as high as that from the shock-compressed
ablator.
The speed of sound in the shocked region is greater than the shock speed,
so the release wave erodes the shock.
The hydrodynamic flow is self-similar, so the erosion of the shock by the
release wave can be characterized by
the angle $\phi$ at which the planar region of the shock is eroded.
$\phi$ is often assumed to be $45^\circ$, but this is not generally the case.
As with the catch-up speed, these angles can be calculated given the sound
speed on the principal Hugoniot $c_a$, 
by considering the propagation of a disturbance from the edge of the planar
shock region, traveling at
$c_a$, across material moving with the particle speed $u_a$ parallel
with the shock, and the resulting speed of the disturbance as it moves across
successive positions of the shock itself:
\begin{eqnarray}
\tan\phi & = & \frac{c_a \sin\phi'}{s_a} \quad:\quad
\cos\phi' = \frac{s_a-u_a}{c_a}. \\
\Rightarrow \tan\phi & = & \frac 1{s_a}\sqrt{c_a^2-\left(s_a-u_a\right)^2}
\end{eqnarray}
An equivalent analysis has been used previously to measure the sound speed
on the Hugoniot \cite{Zeldovich66}.
For a circular laser spot of diameter $d_l$,
the diameter of the region of the sample initially subjected to
the maximum shock pressure is 
\begin{equation}
d_s = d_l - 2 l \tan\phi,
\end{equation}
where $l$ is the thickness of the ablator.
Subsequent release depends also on the sound and particle speed in the 
double-shocked ablator material.
The same analysis, using the Hugoniot state in the sample,
can be used to predict the lateral erosion of the shock in the sample itself.
The edge release angle is important when designing experiments to study
phenomena at the leading edge of the shock.
For experiments on release from the shocked state, including tensile damage,
an important design parameter is the scale rate at which lateral release
propagates radially through the shocked material, $c_a/s_a$.\footnote{%
   The ratio $c_a/s_a$ may be greater or less than one.
   A steady shock is always subsonic with respect to the flow behind.
   However, with respect to the shocked material,
   the shock moves at $s_a-u_a$, which does not exclude $c_a < s_a$.
}

\begin{figure}
\begin{center}\includegraphics[scale=0.70]{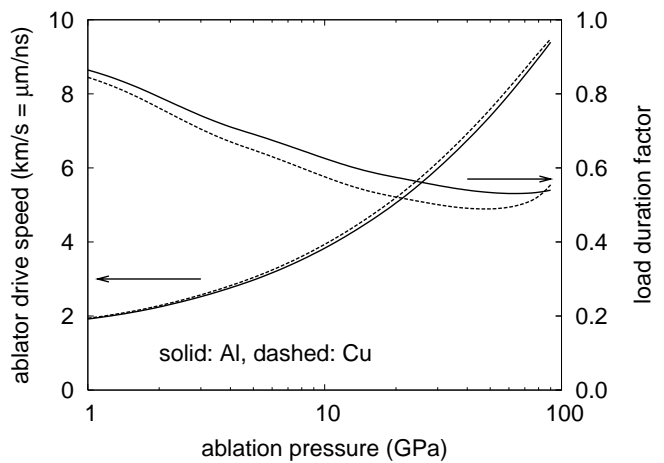}\end{center}
\caption{Example calculations of `ablator drive speed'
   relating the optimum ablator thickness and laser pulse duration,
   and load duration factor relating the laser pulse duration
   and pressure pulse duration on the sample,
   for a parylene-N ablator and Al or Cu samples.}
\label{fig:drivespeeds}
\end{figure}

\begin{figure}
\begin{center}\includegraphics[scale=0.72]{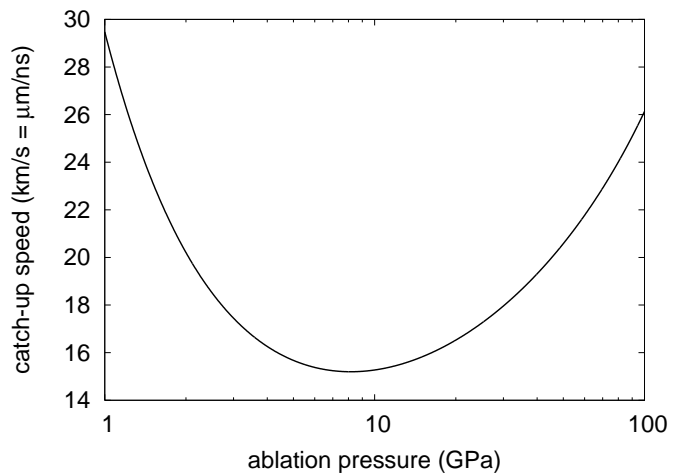}\end{center}
\caption{Example calculations of `catch-up speed'
   relating the maximum ablator thickness and laser pulse duration,
   for a parylene-N ablator.}
\label{fig:catchup}
\end{figure}

\begin{figure}
\begin{center}\includegraphics[scale=0.72]{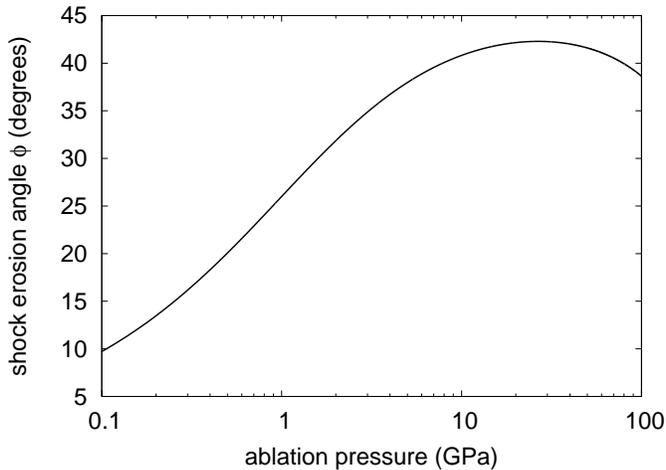}\end{center}
\caption{Shock erosion angle,
   describing the angle at which the lateral release wave
   propagates across the shock, for a parylene-N ablator.}
\label{fig:edgeangle}
\end{figure}

\begin{figure}
\begin{center}\includegraphics[scale=0.72]{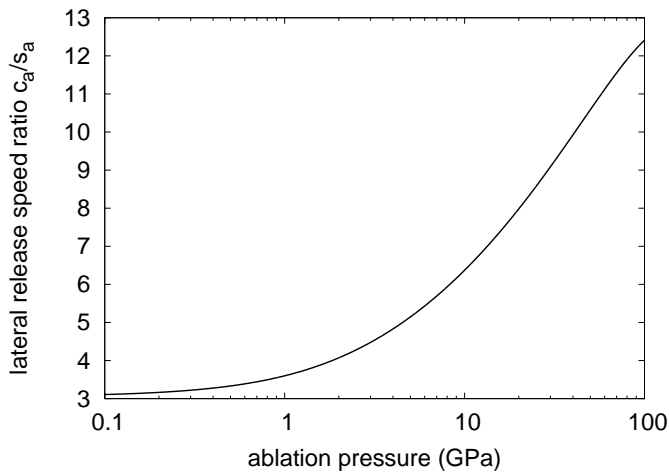}\end{center}
\caption{Ratio between bulk sound speed and shock speed,
   describing the rate at which the lateral release wave
   propagates radially through the shocked materal,
   for a parylene-N ablator.}
\label{fig:latsonicity}
\end{figure}

The ablator drive speed, load duration factor, catch-up speed,
edge angle, and lateral release ratio
were calculated using a generalized shock dynamics algorithm to determine
the principal and secondary Hugoniot states \cite{Swift_genhug07},
using the same EOS as above
(Figs~\ref{fig:drivespeeds} to \ref{fig:latsonicity}).
As was found previously for direct ablation of metals \cite{Swift_maxthick_01},
the catch-up speed exhibited a minimum value around the bulk modulus of the
ablated material: at the minimum, the ablator thickness should be smallest
for a given laser pulse duration.
The ratio $c_a/s_a$ was greater than 3 for all relevant pressures,
and rose past 10 for ablation pressures above 40\,GPa:
lateral release in the ablator may be a concern in experiments with
thick ablators, though the pressure may not drop rapidly once release
starts as it is sustained by the ablation.
Radiation hydrodynamics simulations were performed of the integrated system,
with consistent results for the duration of the peak shock state in the sample.
It should be noted that these quantities depend on material properties
in states significantly off the principal
shock Hugoniot of the ablator.
The hydrodynamic relations above are valid so long as radiation and 
charged particle transport are not significant within the compressed ablator
or the sample.
The radiation hydrodynamics simulations indicated that radiation transport
is unlikely to be significant in this regime; they did not include
electron transport.
For ablatively driven experiments on a range of materials at a range of shock
states, a wide range of off-Hugoniot states may be involved.
These properties are often not known at all accurately from experimental
measurements.
It would be more efficient to use experimental measurements
to validate and adjust
wide-ranging theoretical models, rather than attempting to measure the 
properties of the ablator so generally.
Experiments should include the principal Hugoniot of the ablator,
and also off-Hugoniot states induced by, for example, reflected shocks 
from high-density samples.
These data could be acquired using laser loading experiments with velocimetry
measurements. 
Transparent samples would allow the velocity history to be measured
at the interface.
Whether transparent or opaque, the samples should be thick enough to prevent 
release waves from their free surface from perturbing the states inside
the ablator, which would complicate the experiments.
Measurements would include the amplitude and duration of the shock in the
sample.

\section{Conclusions}
The sensitivity of ablation pressure in CH-based plastic ablators to 
uncertainties in equation of state or composition was predicted to be 5-10\%,
which is small compared with typical uncertainties in laser irradiance.
A laser pulse of constant irradiance was predicted to induce an ablation
pressure that decreased by several percent per nanosecond, but was
significantly more constant than the pressure history induced by direct ablation
of metal samples.
A simple relationship between laser irradiance and shock pressure was
deduced from the simulations, valid for irradiances from 0.1-100\,PW/m$^2$.
Al flashing, used to prevent early-time shine-through, was predicted to make
at most a few percent difference in loading history
once dielectric breakdown had occurred,
increasing the pressure and flattening the pressure history at lower
irradiances.
It was demonstrated that the irradiance history can be adjusted to produce 
a more constant pressure history.

The impedance mismatch between plastic ablators and metal samples
induces a stronger transmitted shock in the sample, by a factor of around 
1.8 for Al and 2.5 for Cu.
The shock reflected into the ablator interacts with the ablation surface 
to produce a stepped pressure history in the sample, the pressure
reducing to closer to the ablation pressure.  This composite loading history
should be taken into account when designing and interpreting material dynamics
experiments.
The shock pressure is applied to the sample for a significantly shorter time 
than the laser pulse duration, because the successive waves in the ablator are
faster than the initial ablation-driven shock.

A compact method was found for representing the optimum ablator thickness for
a given laser pulse duration (or vice versa), as an effective speed.
A similar method was found to represent the relationship between the
laser pulse duration and the duration of the initial, peak pressure pulse
applied to the sample.
These relations are helpful in the design and interpretation of ablatively
driven shock experiments.

\section*{Acknowledgments}
We would like to acknowledge the contributions of
Jeff Colvin for comments and advice on radiation hydrodynamics simulations
and the manuscript,
David Young for providing the equation of state and comments for parylene-N,
and Jon Larson for advice on HYADES simulations,
and Flavien Lambert for comments on the manuscript.
This work was performed in support of
Laboratory-Directed Research and Development project 06-SI-004
(Principal Investigator: Hector Lorenzana),
under the auspices of
the U.S. Department of Energy under contracts
W-7405-ENG-48 and DE-AC52-07NA27344.

\end{document}